\documentclass[reprint,superscriptaddress,amsmath,amssymb,aps,prd]{revtex4-2}
\usepackage{graphicx}
\usepackage{dcolumn}
\usepackage{bm}
\usepackage[mathlines]{lineno}
\usepackage[hidelinks]{hyperref}
\usepackage{orcidlink}
\usepackage{misc/marco}

\newcommand{\cbreview}{\cite{Ni:2007ar,Komatsu:2022nvu}}
\newcommand{\de}{~\cite{Carroll:1998zi,Fujita:2020aqt,Panda:2010uq,Fujita:2020ecn,Choi:2021aze,Obata:2021nql,Gasparotto:2022uqo,Galaverni:2023zhv}~}
\newcommand{\ede}{~\cite{Murai:2022zur,Eskilt:2023nxm,Fujita:2020ecn}~}
\newcommand{\dm}{~\cite{Finelli:2008jv,Liu:2016dcg,Fedderke:2019ajk}~}
\newcommand{\td}{~\cite{Takahashi:2020tqv,Kitajima:2022jzz,Jain:2022jrp,Gonzalez:2022mcx}~}
\newcommand{\qg}{~\cite{Myers:2003fd,Balaji:2003sw,Arvanitaki:2009fg}~}
\newcommand{\recentEB}{~\cite{Minami:2020odp,Diego-Palazuelos:2022dsq,Eskilt:2022cff,Cosmoglobe:2023pgf,ACT:2025fju}}
\newcommand{\ebtomo}{~\cite{Sigl:2018fba,Sherwin:2021vgb,Nakatsuka:2022epj,Liu:2006uh,Lee:2013mqa,Gubitosi:2014cua,Eskilt:2023nxm,Fujita:2020ecn,Finelli:2008jv}}
\newcommand{\ebtomoinst}{~\cite{QUaD:2008ado,Komatsu:2011a,BICEP1:2013rur,Planck:2016soo,Monelli:2022pru}~}
\newcommand{\anisotheory}{~\cite{Lue:1998mq,caldwell:2011a,Lee:2014rpa,Leon:2016kvt,Yin:2023srb,Ferreira:2023jbu}~}

\begin{document}

\title{Constraints on Anisotropic Cosmic Birefringence from CMB B-mode Polarization}

\author{A. I. Lonappan\thanks{ORCID: \href{https://orcid.org/0000-0003-1200-9179}{0000-0003-1200-9179}}}
\email{alonappan@ucsd.edu}
\affiliation{Department of Physics, University of California, San Diego, CA 92093, USA}

\author{B. Keating\thanks{ORCID:\href{https://orcid.org/0000-0003-3118-5514}{0000-0003-3118-5514}}}
\email{bkeating@ucsd.edu}
\affiliation{Department of Physics, University of California, San Diego, CA 92093, USA}

\author{K. Arnold\thanks{ORCID:\href{https://orcid.org/0000-0002-3407-5305}{0000-0002-3407-5305}}}
\email{karnold@ucsd.edu}
\affiliation{Department of Physics, University of California, San Diego, CA 92093, USA}
\affiliation{Department of  Astronomy \& Astrophysics University of California, San Diego, CA 92093, USA}
 
\date{\today}

\begin{abstract}
Cosmic birefringence—the rotation of the polarization plane of light as it traverses the universe—offers a direct observational window into parity-violating physics beyond the Standard Model. In this work, we revisit the anisotropic component of cosmic birefringence, which leads to the generation of $B$-mode polarization in the cosmic microwave background (CMB). Using an exact theoretical treatment beyond the thin last-scattering surface approximation, we constrain the amplitude of anisotropic birefringence with combined polarization data from SPTpol, ACT, POLARBEAR, and BICEP. The joint analysis yields a best-fit amplitude of $A_{\mathrm{CB}} = 0.42^{+0.40}_{-0.34} \times 10^{-4}$, consistent with zero within $2\sigma$, and we place a 95\% confidence-level upper bound of $A_{\mathrm{CB}} < 1 \times 10^{-4}$. The constraint is not dominated by any single experiment and remains robust under the inclusion of a possible isotropic rotation angle. These results provide leading constraints on anisotropic cosmic birefringence from CMB $B$-mode polarization and illustrate the potential of upcoming experiments to improve sensitivity to parity-violating effects in the early universe.
\end{abstract}

\maketitle

\section{Introduction}

Cosmic birefringence\textemdash the rotation of the polarization plane of electromagnetic radiation as it propagates across cosmological distances\textemdash offers a sensitive test of parity-violating physics beyond the Standard Model~\cbreview. This phenomenon can arise when pseudoscalar fields, such as axion-like particles (ALPs), interact with photons through the Chern–Simons coupling, typically represented by an effective Lagrangian term of the form
\begin{equation}
    \mathcal{L} \supset -\frac{g_{\phi}}{4}\phi F_{\mu\nu}\tilde{F}^{\mu\nu}.
\end{equation} 
Here, $g_{\phi}$ is the coupling constant, $\phi$ denotes the pseudoscalar field, $F_{\mu\nu}$ is the electromagnetic tensor, and $\tilde{F}^{\mu\nu}$ is its dual. Various cosmological scenarios, including dark matter\dm, early dark energy\ede, dark energy\de and topological defects\td, have been proposed as potential sources for ALPs driving cosmic birefringence. Moreover, quantum gravity models also predict detectable signatures through similar parity-violating effects\qg.

Recent analyses of CMB polarization data have uncovered suggestive signals that may indicate cosmic birefringence~\cite{ACT:2025fju}. The cross-correlations between even-parity $E$-modes and odd-parity $B$-modes, from Planck polarization maps, have provided a provisional hint for this effect, with reported rotation angles achieving moderate statistical significance\recentEB, spurring interest in more detailed investigations. Importantly, the time evolution of pseudoscalar fields during the recombination and reionization epochs can significantly impact the $EB$ power spectrum, modifying both its amplitude and spectral shape. Precise characterization of this spectral structure thus offers a valuable tomographic probe of pseudoscalar field dynamics across cosmic history\ebtomo. Additionally, such a tomographic analysis helps to mitigate degeneracies associated with instrumental systematics\ebtomoinst.

Fluctuations in pseudoscalar fields induce spatially varying rotations of the polarization plane, thereby resulting in anisotropic cosmic birefringence \anisotheory. Importantly, anisotropic cosmic birefringence induces mode coupling within the polarization fields, enabling reconstruction of the birefringence angle through quadratic estimator techniques applied to CMB polarization maps~\cite{Gluscevic:2009mm}. To date, anisotropic birefringence has not been conclusively detected, and current observations have yielded only upper limits on its amplitude~\cite{BICEP2:2017lpa,Namikawa:2020ffr}. Planned advancements in CMB polarization experiments, including BICEP~\cite{Moncelsi:2020ppj}, Simons Observatory~\cite{SimonsObservatory:2018koc}, CMB-S4~\cite{CMB-S4:2020lpa}, and LiteBIRD~\cite{LiteBIRD:2022cnt} are anticipated to significantly reduce both instrumental noise and systematic uncertainties. These improvements will enhance the detectability of cosmic birefringence signals, facilitating tighter constraints on both isotropic and anisotropic birefringence and thus offering a powerful probe into new physics scenarios~\cite{IdicherianLonappan:2025trj}.

Similar to the B-modes induced by isotropic birefringence, anisotropic birefringence also converts E-mode polarization into B-mode polarization. Previous analyses of anisotropic birefringence typically employed the approximation of a thin last-scattering surface, neglecting the time evolution of the pseudoscalar fields during recombination and reionization to simplify the calculations~\cite{Liu:2016dcg}. However, recent studies, notably by T.~Namikawa~\cite{Namikawa:2024dgj} (hereafter TN24), have highlighted the importance of accurately accounting for this finite thickness and the corresponding evolution of the pseudoscalar fields. TN24 derived and numerically computed the exact B-mode power spectrum arising from anisotropic cosmic birefringence without relying on the thin LSS approximation. Building upon this theoretical foundation, we update the constraints presented in TN24 by incorporating additional polarization measurements from the Atacama Cosmology Telescope (ACT), POLARBEAR, and BICEP experiments, aiming to provide more robust limits on the amplitude of anisotropic birefringence.

This paper is organized as follows. In Sec.~\ref{sec:theory}, we review the exact computation of the $B$-mode power spectrum sourced by anisotropic cosmic birefringence. Sec.~\ref{sec:likelihood} presents the likelihood analysis used to constrain the birefringence amplitude. We summarize our findings and discuss future prospects in Sec.~\ref{sec:conclusion}. 
\section{Theoretical Framework}
\label{sec:theory}
This section presents the theoretical framework for computing the CMB B-mode power spectrum induced by anisotropic cosmic birefringence, following the exact treatment of TN24, which accounts for the finite thickness of the last-scattering surface (LSS) rather than assuming a thin LSS. We summarize TN24’s methodology, provide key equations, and discuss the conditions under which the approach is valid.
\subsection{Rotation Angle and Pseudoscalar Field Evolution}
For a massless pseudoscalar field with spatial fluctuations $\delta \phi$, the anisotropic rotation angle of CMB photon polarization along a line-of-sight direction $\hat{\boldsymbol{n}}$ at conformal time $\eta$ is given by~\citep{Harari:1992ea,Carroll:1989vb,PhysRevD.43.3789}:
\begin{equation}
\alpha(\eta, \hat{\boldsymbol{n}}) = -\frac{g_\phi}{2} \delta \phi(\eta, \chi \hat{\boldsymbol{n}}),
\label{eq:rotation_angle}
\end{equation}
where $\chi = \eta_0 - \eta$ is the comoving distance, $\eta_0$ is the present conformal time, and $g_\phi$ is the coupling constant introduced in the Chern-Simons term. The isotropic rotation component vanishes for a massless field. The evolution of the pseudoscalar field fluctuations is governed by:
\begin{equation}
\frac{\mathrm{d}^2 \delta \phi}{\mathrm{d} \eta^2} + 2 \mathcal{H} \frac{\mathrm{d} \delta \phi}{\mathrm{d} \eta} + k^2 \delta \phi = 0,
\label{eq:phi_evolution}
\end{equation}
where $\mathcal{H} = (\mathrm{d} a / \mathrm{d} \eta) / a$ is the conformal Hubble parameter, and $a$ is the scale factor. The fluctuations are described by a transfer function:
\begin{equation}
\delta \phi(\eta, \boldsymbol{k}) = T(k, \eta) \delta \phi_{\text{ini}}(\boldsymbol{k}),
\label{eq:transfer_function}
\end{equation}
with $T(k, \eta) = 3 \frac{j_1(k \eta)}{k \eta}$ during matter domination, where $j_1$ is the spherical Bessel function of the first kind. The primordial power spectrum, arising from vacuum fluctuations during inflation, is:
\begin{equation}
\langle \delta \phi_{\text{ini}}^*(\boldsymbol{k}) \delta \phi_{\text{ini}}(\boldsymbol{k}') \rangle = \frac{2 \pi^2}{k^3} \mathcal{P}_\phi(k) (2 \pi)^3 \delta^{(3)}(\boldsymbol{k} - \boldsymbol{k}'),
\label{eq:primordial_power}
\end{equation}
where \(\mathcal{P}_\phi(k) = \left( \frac{H_I}{2 \pi} \right)^2\), and \(H_I\) is the Hubble parameter during inflation~\citep{Lue:1998mq,Greco:2022xwj}.
The rotation modifies the Stokes parameters as:
\begin{equation}
[Q \pm \mathrm{i} U](\hat{\boldsymbol{n}}) = [\bar{Q} \pm \mathrm{i} \bar{U}](\hat{\boldsymbol{n}}) \mathrm{e}^{\pm 2 \mathrm{i} \alpha(\eta, \hat{\boldsymbol{n}})},
\label{eq:stokes_rotation}
\end{equation}
where \(\bar{Q}\) and \(\bar{U}\) are the unrotated Stokes parameters. The E- and B-mode polarization fields are defined using spin-2 spherical harmonics:
\begin{equation}
E_{\ell m} \pm \mathrm{i} B_{\ell m} = -\int \mathrm{d}^2 \hat{\boldsymbol{n}} \, {}_{\pm 2} Y_{\ell m}^*(\hat{\boldsymbol{n}}) [Q \pm \mathrm{i} U](\hat{\boldsymbol{n}}).
\label{eq:eb_modes}
\end{equation}
\subsection{B-Mode Power Spectrum Without Thin LSS Approximation}
Earlier studies often approximated the LSS as infinitely thin, fixing the rotation angle at a single conformal time $\eta_*$ (e.g.,~\citep{Li:2008tma}). This simplification neglects the time evolution of $\delta \phi$ during recombination and reionization, overestimating the B-mode signal. TN24 employs the total angular momentum method~\citep{Hu:1997hp} to derive an exact B-mode power spectrum, accounting for the finite LSS thickness~\citep{Pogosian:2011qv}.
The evolution of the polarization fields \(P_{\pm} = Q \pm \mathrm{i} U\) in Fourier space is given by~\citep{Pogosian:2011qv,Nakatsuka:2022epj}:
\begin{align}
\frac{\mathrm{d} P_{\pm}}{\mathrm{d} \eta}(\eta, \boldsymbol{q}, \hat{\boldsymbol{n}})
&+ \mathrm{i} q \mu\, P_{\pm}(\eta, \boldsymbol{q}, \hat{\boldsymbol{n}}) \notag \\
&= \dot{\tau} \bigg[ 
    -P_{\pm}(\eta, \boldsymbol{q}, \hat{\boldsymbol{n}})
    + \sqrt{6}\, P^{(0)}(\eta, \boldsymbol{q}) \notag \\
&\quad \times \sqrt{\frac{4\pi}{5}}\, {}_{\pm 2}Y_{20}(\hat{\boldsymbol{n}})
  \bigg] \notag \\
&\quad \mp 2 \mathrm{i} \frac{\mathrm{d} \alpha}{\mathrm{d} \eta}(\eta, \hat{\boldsymbol{n}})
    P_{\pm}(\eta, \boldsymbol{q}, \hat{\boldsymbol{n}}),
\label{eq:polarization_evolution}
\end{align}
where $\mu = \hat{\boldsymbol{n}} \cdot \hat{\boldsymbol{q}}$, $\dot{\tau} = a n_e \sigma_T$ is the differential CMB optical depth, $n_e$ is the electron number density, $\sigma_T$ is the Thomson scattering cross-section, and $P^{(0)} = (\Theta_2 - \sqrt{6} E_2) / 10$ is the polarization source from scalar perturbations~\citep{Hu:1997hp}. Solving this equation, the polarization fields are:
\begin{align}
P_{\pm}(\boldsymbol{q}, \hat{\boldsymbol{n}}) 
&= -\int_0^{\eta_0} \mathrm{d} \eta \, g_v(\eta) \sqrt{6}\, P^{(0)}(\eta, \boldsymbol{q}) 
    [1 \pm 2 \mathrm{i} \alpha(\eta, \hat{\boldsymbol{n}})] \notag \\
&\quad \times \sum_{\ell} (-\mathrm{i})^{\ell} \sqrt{4\pi (2\ell + 1)}\, 
    \epsilon_{\ell}^{(0)}(q (\eta_0 - \eta)) \,
    {}_{\pm 2}Y_{\ell 0}(\hat{\boldsymbol{n}}) \notag \\
&\quad + \mathcal{O}(\alpha^2),
\label{eq:polarization_solution}
\end{align}
where \(g_v(\eta)\) is the visibility function, and \(\epsilon_{\ell}^{(0)}(x) = \sqrt{\frac{3}{8} \frac{(\ell + 2)!}{(\ell - 2)!}} \frac{j_\ell(x)}{x^2}\). Decomposing \(P_{\pm}\) into E- and B-modes and computing the power spectrum, TN24 derives the B-mode power spectrum as:
\begin{equation}
C_{\ell}^{BB} = 4 \sum_{\ell' L} p_{\ell L \ell'}^{+} \frac{(2 \ell' + 1)(2 L + 1)}{4 \pi} \left( \begin{array}{ccc} \ell & L & \ell' \\ 2 & 0 & -2 \end{array} \right)^2 \mathcal{C}_{\ell', L}^{EE},
\label{eq:bb_power_spectrum}
\end{equation}
where \(p_{\ell L \ell'}^{+} = [1 - (-1)^{\ell + L + \ell'}] / 2\) enforces parity symmetry, and the distorted E-mode power spectrum is:
\begin{equation}
\mathcal{C}_{\ell', L}^{EE} = 4 \pi \int \mathrm{d} \ln k \, \mathcal{P}_\phi(k) \mathcal{C}_{\ell', L}^{EE}(k),
\label{eq:ee_distorted}
\end{equation}
with:
\begin{equation}
\mathcal{C}_{\ell', L}^{EE}(k) = 4 \pi \int \mathrm{d} \ln q \, \mathcal{P}_{\mathcal{R}}(q) \left[ \Delta_{\ell', L}(q, k) \right]^2,
\label{eq:ee_k}
\end{equation}
and:
\begin{equation}
\Delta_{\ell', L}(q, k) = \int_0^{\eta_0} \mathrm{d} \eta \, s_{\ell'}(q, \eta) u_L(k, \eta).
\label{eq:delta}
\end{equation}
Here, \(\mathcal{P}_{\mathcal{R}}(q)\) is the primordial curvature power spectrum, \(s_{\ell}(q, \eta) = g_v(\eta) \sqrt{6} S(\eta, q) \epsilon_{\ell}^{(0)}(q (\eta_0 - \eta))\) is the projected polarization source, and \(u_L(k, \eta) = \frac{g_\phi}{2} j_L(k (\eta_0 - \eta)) T(k, \eta)\) encodes the birefringence angle’s evolution. The angular power spectrum of the rotation angle is:
\begin{equation}
C_L^{\alpha \alpha}(\eta, \eta') = 4 \pi \int \mathrm{d} \ln k \, \mathcal{P}_\phi(k) u_L(k, \eta) u_L(k, \eta').
\label{eq:alpha_power}
\end{equation}
At large angular scales (\(L \lesssim 100\)), the rotation angle power spectrum scales as \(C_L^{\alpha \alpha} \simeq \frac{2 \pi}{L (L + 1)} A_{\text{CB}}\), where \(A_{\text{CB}} = \left( \frac{g_\phi}{2} \right)^2 \left( \frac{H_I}{2 \pi} \right)^2\) parametrizes the birefringence amplitude~\citep{2011PhRvD..84d3504C}.
\subsection{TN24 Methodology and Validity}
TN24 implements the above formalism numerically using a modified version of the CLASS code~\citep{Diego_Blas_2011}, computing \(\mathcal{C}_{\ell', L}^{EE}(k)\) and integrating over wavenumbers \(k\) and \(q\) to obtain \(C_{\ell}^{BB}\). The exact treatment captures the time evolution of \(\delta \phi\), resulting in a B-mode power spectrum suppressed by approximately an order of magnitude at large scales (\(\ell \lesssim 10\)) and a factor of two at smaller scales (\(\ell \gtrsim 100\)) compared to the thin LSS approximation.
The TN24 methodology assumes:
\begin{itemize}
    \item A massless pseudoscalar field, ensuring no correlation between \(\delta \phi\) and curvature perturbations, which simplifies the power spectrum calculation.
    \item Negligible gravitational lensing effects on the birefringence angle, justified for a steep red spectrum where lensing contributions are subdominant.
    \item Higher-order terms (\(\mathcal{O}[(C_L^{\alpha \alpha})^2]\)) and lensing contributions to the B-mode power spectrum are minor for \(\ell \lesssim 1000\).
\end{itemize}
These assumptions are valid for scale-invariant anisotropic birefringence spectra and models where the pseudoscalar field’s evolution during recombination and reionization significantly impacts the polarization signal. Our analysis adopts TN24’s framework, utilizing the publicly available \texttt{biref-aniso-bb}\footnote{https://github.com/toshiyan/biref-aniso-bb} code to compute \(C_{\ell}^{BB}\) for our likelihood analysis.
\begin{figure}[h!]
    \centering
    \includegraphics[width=0.9\linewidth]{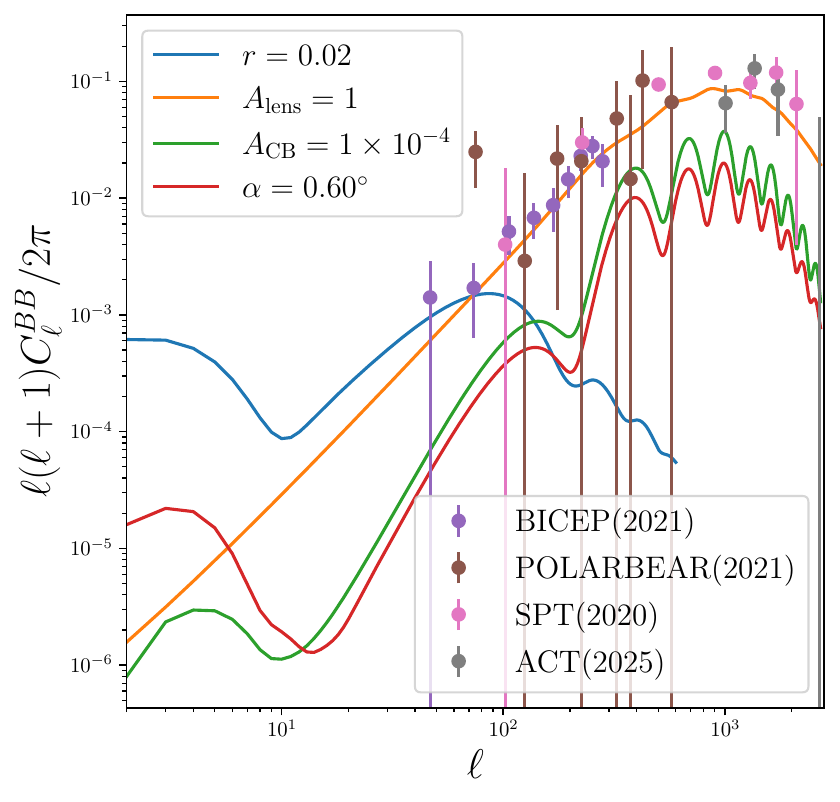}
\caption{Theoretical B-mode spectra are shown for tensor modes with $r = 0.02$ (blue), 
lensing with $A_{\mathrm{lens}} = 1$ (orange), anisotropic birefringence with 
$A_{\mathrm{CB}} = 1 \times 10^{-4}$ (green) and isotropic rotation with $\alpha = 0.60^\circ$ (red). Overlaid error bars are the measurements from BICEP (purple), POLARBEAR (brown), SPTpol (pink), and ACT (grey).}
    \label{fig:bb_spectra}
\end{figure}
Figure~\ref{fig:bb_spectra} shows the theoretical B-mode spectra for a tensor component 
with $r=0.02$, a lensing contribution with $A_{\mathrm{lens}}=1$, an anisotropic 
birefringence contribution with $A_{\mathrm{CB}}=1\times 10^{-4}$ and an isotropic rotation contribution with an angle $\alpha =0.60^\circ$. The measured spectra 
from BICEP~\cite{BICEP:2021xfz}, POLARBEAR~\cite{POLARBEAR:2022dxa}, SPTpol~\cite{SPT:2019nip}, and ACT~\cite{ACT:2025fju} are shown as colored points with error bars.
\section{Likelihood analysis}\label{sec:likelihood}
In the following, we describe the model used specifically in the SPTpol analysis. This treatment differs from how other datasets are incorporated, as we perform a detailed component separation and parameter fitting only for SPTpol, while the other datasets enter through foreground-marginalized combined spectra. We fit the cross-frequency $B$-mode bandpowers of SPTpol using the model:
\begin{align}
D_\ell^{\nu_i \times \nu_j} &=\ 
r\, D_\ell^{\mathrm{tens}} 
+ A_{\mathrm{lens}}\, D_\ell^{\mathrm{lens}} \notag \\
&\quad + A_{\mathrm{CB}}\, D_\ell^{\mathrm{aniso,CB}} 
+ D_\ell^{\mathrm{iso,CB}}(\alpha) 
+ D_\ell^{\mathrm{fg};\,\nu_i \times \nu_j},
\end{align}
where $D_\ell^{\mathrm{tens}}$ and $D_\ell^{\mathrm{lens}}$ are templates for the tensor and lensing contributions, respectively. The isotropic CB rotation is modeled as
\begin{equation}
D_\ell^{\mathrm{iso,CB}}(\alpha) = \sin(2\alpha)\,C_\ell^{\mathrm{EE}} 
\end{equation}
where $\alpha$ is the global birefringence angle (or a polarization angle miscalibration), and $C_\ell^{\mathrm{EE}}$, $C_\ell^{\mathrm{BB}}$ are the lensed $E$- and $B$-mode power spectra. All theoretical templates are computed using \texttt{CAMB}\footnote{\url{https://github.com/cmbant/CAMB}} with the \textit{Planck} 2018 best-fit cosmology. The B-mode spectrum from anisotropic birefringence, $D_\ell^{\mathrm{aniso,CB}}$, is computed using the \texttt{biref-aniso-bb} code\footnote{\url{https://github.com/toshiyan/biref-aniso-bb}}.

The foreground term is modeled as:
\begin{align}
D_\ell^{\mathrm{fg};\,\nu_i \times \nu_j} &=\ 
A_{\ell=80}^{\mathrm{dust};\,150\,\mathrm{GHz}}\, f_{\nu_i \nu_j}
\left( \frac{\ell}{80} \right)^{-0.58} \notag \\
&\quad + A_{\ell=3000}^{\mathrm{Pois};\,\nu_i \times \nu_j} 
\left( \frac{\ell}{3000} \right)^2,
\end{align}
where $f_{\nu_i \nu_j}$ captures the frequency scaling of Galactic dust, modeled as a modified blackbody with temperature $T = 19.6\,\mathrm{K}$ and spectral index $\beta = 1.59$. We impose a Gaussian prior on the dust amplitude, $A_{\ell=80}^{\mathrm{dust};\,150\,\mathrm{GHz}} = 0.0094 \pm 0.0021\,\mu\mathrm{K}^2$, based on BICEP2/Keck measurements, and also apply a Gaussian prior on $A_{\mathrm{lens}}$. Flat priors are used for the birefringence amplitude $A_{\mathrm{CB}}$, the isotropic CB angle $\alpha$, and the tensor-to-scalar ratio $r$.

Following the SPTpol analysis, the model includes two calibration parameters (one for each frequency band), and seven nuisance parameters to marginalize over uncertainties in the beam window functions. When incorporating external data from ACT, POLARBEAR, and BICEP, we use combined spectra that are already marginalized over foreground contributions, and in this case, we fit only for $A_{\mathrm{CB}}$, $\alpha$, and $r$.

The likelihood is assumed to be Gaussian in the bandpowers:
\begin{multline}
-2 \ln \mathcal{L}(\boldsymbol{\theta}) = 
\sum_{b, b'} 
\left( D_b^{\mathrm{obs}} - D_b^{\mathrm{model}}(\boldsymbol{\theta}) \right)
\left( \mathbf{C}^{-1} \right)_{b b'} \\
\times 
\left( D_{b'}^{\mathrm{obs}} - D_{b'}^{\mathrm{model}}(\boldsymbol{\theta}) \right),
\end{multline}
where $D_b^{\mathrm{obs}}$ are the observed B-mode bandpowers, 
$D_b^{\mathrm{model}}(\boldsymbol{\theta})$ is the theoretical prediction 
as described above, and $\mathbf{C}$ is the covariance matrix of the data. 
The parameter vector $\boldsymbol{\theta}$ includes the cosmological parameters 
($r$, $A_{\mathrm{lens}}$, $A_{\mathrm{CB}}$, $\alpha$), foreground amplitudes, 
calibration parameters, and beam nuisance parameters.

We sample the posterior distribution using the \texttt{emcee} 
affine-invariant Markov Chain Monte Carlo (MCMC) sampler~\cite{emcee}, 
marginalizing over all nuisance parameters and priors described above.\\
\begin{table*}[t!]
    \centering
    \caption{Constraints on cosmic birefringence amplitude $A_{\rm CB}$ in units of $\times 10^{-4}$, along with detection significance and the 95\% confidence level upper limits for each dataset combination.}
    \label{tab:Acb_constraints}
    \begin{tabular}{lccc}
        \hline
        \textbf{Model} & $A_{\rm CB}$ [$\times 10^{-4}$] & Significance & 95\% CL Upper Limit [$\times 10^{-4}$] \\
        \hline
        SPTpol & $0.97^{+0.55}_{-0.52}$ & $1.80\sigma$ & 1.88 \\
        SPTpol + ACT & $0.51^{+0.37}_{-0.43}$ & $1.27\sigma$ & 1.12 \\
        SPTpol + ACT + POLARBEAR & $0.49^{+0.39}_{-0.41}$ & $1.24\sigma$ & 1.13 \\
        SPTpol + ACT + BICEP & $0.41^{+0.40}_{-0.35}$ & $1.05\sigma$ & 1.07 \\
        SPTpol + ACT + POLARBEAR + BICEP & $0.42^{+0.40}_{-0.34}$ & $1.10\sigma$ & 1.08 \\
        ACT + POLARBEAR + BICEP & $0.01^{+0.51}_{-0.00}$ & $0.02\sigma$ & 0.85 \\
        \hline
    \end{tabular}
\end{table*}

\begin{figure}[h!]
    \centering
    \includegraphics[width=0.9\linewidth]{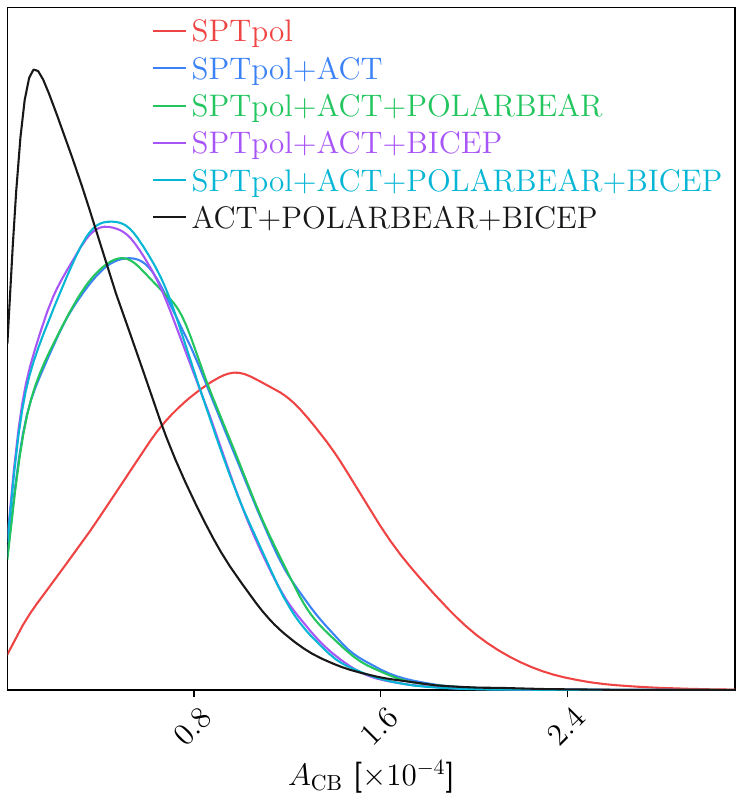}
\caption{Posterior probability distributions for the amplitude $A_{\mathrm{CB}}$ of anisotropic cosmic birefringence, sampled from different dataset combinations (SPTpol, ACT, POLARBEAR, and BICEP).}
    \label{fig:acb_pos}
\end{figure}
To isolate the anisotropic birefringence contribution, we did not include the isotropic term $D_\ell^{\mathrm{iso,CB}}(\alpha)$ in our model. Table~\ref{tab:Acb_constraints} summarizes the updated constraints on the amplitude of anisotropic cosmic birefringence, $A_{\rm CB}$, expressed in units of $10^{-4}$, derived from various combinations of CMB polarization datasets: SPTpol, ACT, POLARBEAR, and BICEP. For each case, the table reports the best-fit value with 68\% confidence intervals and the corresponding detection significance. The SPTpol-only result yields $A_{\rm CB} = 0.97^{+0.55}_{-0.52} \times 10^{-4}$. When ACT data are included, the preferred amplitude shifts to $0.51^{+0.37}_{-0.43} \times 10^{-4}$. Adding POLARBEAR and BICEP further reduces the best-fit amplitude, with the full combination yielding $A_{\rm CB} = 0.42^{+0.40}_{-0.34} \times 10^{-4}$ at $1.1\sigma$. This result is consistent with the null hypothesis at the $2\sigma$ level. The dataset combination excluding SPTpol yields a 95\% confidence-level upper limit of $A_{\rm CB} < 1.00 \times 10^{-4}$, and the 68\% interval from the full-dataset result lies entirely within this bound, indicating consistency. Figure~\ref{fig:acb_pos} illustrates the shift in the posterior distributions for $A_{\rm CB}$ as additional datasets are incorporated, reflecting the decreasing amplitude preference and compatibility across experiments.
\begin{figure}[h!]
    \centering
    \includegraphics[width=\linewidth]{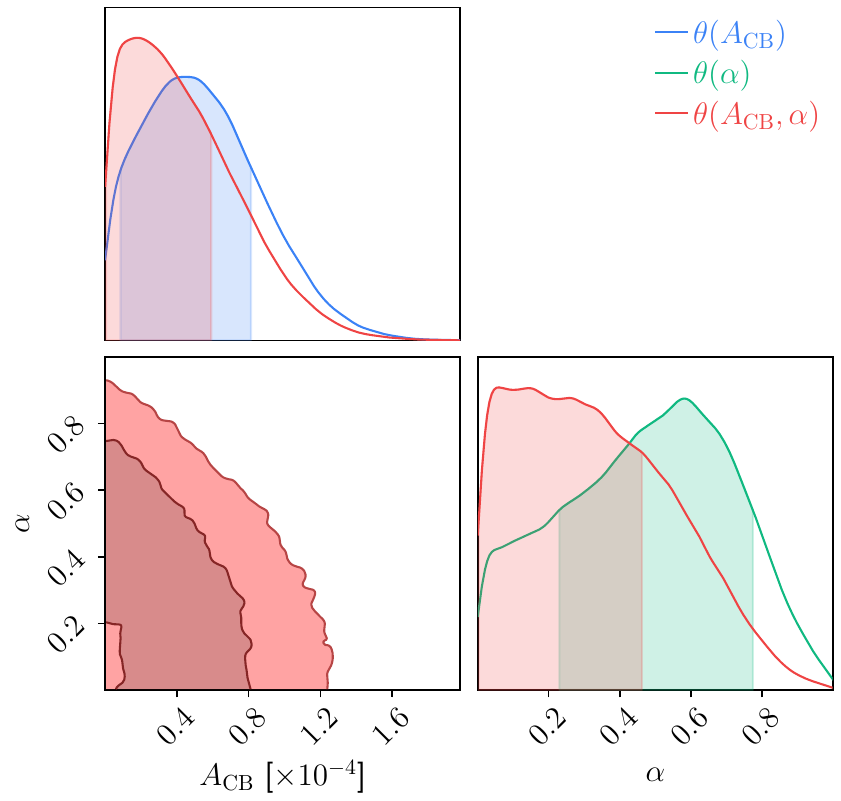}
    \caption{Corner plot showing the marginalized and joint posterior distributions for the anisotropic cosmic birefringence amplitude $A_{\rm CB}$ and the isotropic rotation angle $\alpha$ from the combined dataset. The blue curve represents the marginalized constraint on $A_{\rm CB}$ alone, yielding a best-fit of $4.6^{+3.5}_{-3.8} \times 10^{-5}$. The green curve shows the marginalized constraint on $\alpha$ alone, centered at $0.58^{+0.19}_{-0.35}$ degrees. The red contours correspond to the joint fit, $\theta(A_{\rm CB},\alpha)$, where parameter degeneracy broadens the posteriors and weakens the constraint. From this joint fit, we find no statistically significant detection of $A_{\rm CB}$, and report a 95\% confidence level upper limit of $A_{\rm CB} < 1.0 \times 10^{-4}$. The shaded region represents the 68\% confidence interval.}
    \label{fig:acb_dif}
\end{figure}
\subsection{Isotropic Rotation}\label{subsec:isotropic}
To account for a possible instrumental miscalibration or an isotropic component of cosmic birefringence, we extend our model to include the isotropic rotation term $D_\ell^{\mathrm{iso,CB}}(\alpha)$, where $\alpha$ denotes a constant rotation angle across the sky. This term captures contributions from both physical isotropic birefringence and systematic uncertainties in the absolute polarization angle. We perform a joint likelihood analysis by simultaneously sampling the anisotropic amplitude $A_{\mathrm{CB}}$ and the isotropic angle $\alpha$, and compare the result to the case where only $A_{\mathrm{CB}}$ is varied. Including the isotropic angle shifts the best-fit amplitude of anisotropic birefringence from $(4.6^{+3.5}_{-3.8}) \times 10^{-5}$ in the single-parameter fit to $(1.9^{+4.1}_{-1.8}) \times 10^{-5}$ when both parameters are varied. As shown in Fig.~\ref{fig:acb_dif}, the posterior distribution becomes broader and peaks closer to zero, reflecting a degeneracy between $\alpha$ and $A_{\mathrm{CB}}$, where the isotropic component partially absorbs signal power that might otherwise be interpreted as anisotropic. While the constraint remains consistent with a nonzero value, the inclusion of $\alpha$ reduces the inferred amplitude and weakens the apparent preference for anisotropic birefringence. This highlights the importance of jointly modeling isotropic and anisotropic contributions to obtain unbiased and robust constraints on cosmic birefringence.

Alternative mechanisms, such as Faraday rotation from primordial magnetic fields with $\nu^{-2}$ frequency scaling~\citep{Kosowsky:2004zh}, Lorentz-symmetry violation with a distinct angular imprint~\citep{Kostelecky:2007zz}, and mass-dependent axion-like particles that produce scale dependent spectra~\citep{Greco2022Probing}, can mimic cosmic birefringence. Various scalar-field models also yield similar spectra~\citep{Yin:2023srb}. These effects lie beyond the scope of the present work, which adopts TN24’s scale-invariant, massless pseudoscalar model. We will explore these alternative scenarios in future studies.

\section{Discussion and Conclusion}\label{sec:conclusion}

In this work, we revisited the impact of anisotropic cosmic birefringence on the CMB $B$-mode polarization power spectrum, building on the theoretical formalism established in TN24. Using the exact treatment of birefringence-induced $B$-modes without the thin last-scattering surface approximation, we updated the constraints on the birefringence amplitude $A_{\mathrm{CB}}$ by incorporating recent polarization measurements from ACT, POLARBEAR, and BICEP, alongside the previously used SPTpol data. These additional datasets provide complementary multipole coverage and independent instrumental characteristics, enabling cross-validation and improved control of systematics.

Our updated analysis finds that the $\sim2\sigma$ preference for nonzero anisotropic birefringence observed in the SPTpol-only case weakens when combined with ACT, POLARBEAR, and BICEP. The full combination yields $A_{\mathrm{CB}} = 0.42^{+0.40}_{-0.34} \times 10^{-4}$, which remains consistent with a zero amplitude. The dataset combination excluding SPTpol results in a 95\% confidence-level upper limit of $A_{\mathrm{CB}} < 1.00 \times 10^{-4}$, and the full-dataset result lies entirely within this bound, suggesting consistency across experiments.

We further examined the effect of an isotropic rotation component by jointly sampling $A_{\mathrm{CB}}$ and an overall isotropic angle $\alpha$. Including $\alpha$ shifts the best-fit amplitude from $(4.6^{+3.5}_{-3.8}) \times 10^{-5}$ to $(1.9^{+4.1}_{-1.8}) \times 10^{-5}$, and the posterior distribution peaks closer to zero. This behavior reflects a degeneracy between isotropic and anisotropic contributions: when isotropic rotation is allowed to vary, it can partially absorb the signal attributed to anisotropic birefringence. Although the constraint remains compatible with a nonzero value, the inclusion of $\alpha$ reduces the inferred amplitude and weakens the preference, underscoring the importance of modeling both components simultaneously to avoid overestimating the anisotropic signal.

Looking ahead, upcoming experiments such as Simons Observatory, CMB-S4, and LiteBIRD are expected to deliver significantly improved polarization sensitivity and tighter control over angle calibration. These capabilities will help disentangle isotropic and anisotropic birefringence and enable more definitive searches for parity-violating physics in the early universe. Until then, conservative approaches that marginalize over isotropic rotation, as employed here, remain essential for deriving robust and unbiased constraints. The analysis code used in this work is publicly available at \url{https://github.com/antolonappan/bbCAB}.

\begin{acknowledgments}
The authors thank Toshiya Namikawa for valuable discussions and comments that helped improve both the analysis and interpretation of the results.
\end{acknowledgments}


\bibliography{misc/ref}

\end{document}